\documentclass[preprint,aps,pra,showpacs,floatfix]{revtex4-1}
\usepackage{times}
\usepackage {euscript}
\usepackage{amsthm}
\usepackage{graphicx}
\usepackage{amsmath}
\usepackage{amssymb}
\usepackage{amsfonts}
\usepackage[english]{babel}
\usepackage{epstopdf}
\usepackage{multirow,makecell,array}
\graphicspath{{figures/}}

\begin{document}
\thispagestyle{empty}

    \title{Nuclear recoil and vacuum-polarization effects\\on the binding energies of supercritical H-like ions}

\author{I.~A.~Aleksandrov$^{1, 2}$, G. Plunien$^{3}$, and V.~M.~Shabaev$^{1}$}

\affiliation{$^1$~Department of Physics, St. Petersburg State University, Ulianovskaya 1, Petrodvorets, 198504 Saint Petersburg, Russia\\$^2$~ITMO University, Kronverkskii ave 49, 197101 Saint Petersburg, Russia\\$^3$~Institut f\"ur Theoretische Physik, TU Dresden, Mommsenstrasse 13, Dresden, D-01062, Germany
\vspace{10mm}
}

\begin{abstract}
The Dirac Hamiltonian including nuclear recoil and vacuum-polarization operators is considered in a supercritical regime $Z > 137$. It is found that the nuclear recoil operator derived within the Breit approximation ``regularizes'' the Hamiltonian for the point-nucleus model and allows the ground state level to go continuously down and reach the negative energy continuum at a critical value $Z_\text{cr} \approx 145$. If the Hamiltonian contains both the recoil operator and the Uehling potential, the $1s$ level reaches the negative energy continuum at $Z_\text{cr} \approx 144$. The corresponding calculations for the excited states have been also performed. This study shows that, in contrast to previous investigations, a point-like nucleus can have effectively the charge $Z > 137$.

\end{abstract}

\maketitle
\section{Introduction}
\label{sec:intro}
A great amount of theoretical investigations has been devoted to the problem of the electronic bound states in hydrogenlike ions. For the case of a point nucleus the Dirac equation leads to the well-known Sommerfeld formula which describes the electronic spectrum for $Z < 137$ ($Z$ is the charge number of the nucleus). When $Z$ becomes greater than $137$, the corresponding value of the $1s$ state energy is no longer real. However, the model of an extended nucleus~\cite{gershtein_zeldovich, pieper_greiner, zeldovich_popov, greiner_qed_sf}, which is more realistic, helps one to avoid such intricacy. In this case the energy keeps decreasing and the state ``dives'' into the negative energy continuum at a certain value of $Z$. In Ref.~\cite{gaertner_1981} this phenomenon was investigated with regard to the concept of the vacuum charge. It turned out that when the radius of a supercritical nucleus ($Z > 137$) tends to zero, the vacuum charge screens the nuclear charge to $137$ that prevents a further diving of the electron states. This result leads to the conclusion that the interaction with a point charge in quantum electrodynamics cannot effectively have the coupling strength greater than $1$~\cite{greiner_qed_sf, gaertner_1981}. The present work basically aims at examination how the previous statement may alter in the presence of nuclear recoil and vacuum-polarization operators.

\indent We will consider the full nuclear recoil operator derived within the Breit approximation and investigate the contribution of the vacuum-polarization effect which is described in the leading order by the Uehling potential. The corresponding numerical procedures have been developed and the results can be found in Section~\ref{sec:results}.

\indent Although in this paper the problem is mostly considered from the physical point of view, it is worth noting that its mathematical aspects have been extensively discussed by other authors~\cite{case_1950, burnap_1981, xia_1999, hogreve_2013, gitman_2013}. Despite the fact that a rigorous and consistent theory of self-adjoint operators was applied to the problem, the electronic states cannot be still completely determined for $Z > 137$. There is an ambiguity which does not allow one to choose the ``real'' solution. We provide a brief mathematical discussion of this problem in Section~\ref{sec:math}.

\indent We employ the relativistic units ($\hbar = c = 1$) and the Heaviside charge unit ($\alpha = e^2/4\pi$) throughout the paper.
%%%
\section{Dirac Hamiltonian including nuclear recoil and vacuum-polarization operators}
\label{sec:theory}
First, we consider the case of a point nucleus and the simple Dirac Hamiltonian which neither includes the nuclear recoil nor the vacuum-polarization operator. The Dirac equation in coordinate space:
\begin{equation}
\Big [ \boldsymbol{\alpha} \cdot \boldsymbol{p} + \beta m + V_\text{C} \Big ] \Psi (\boldsymbol{r}) = E\Psi (\boldsymbol{r}),
\label{eq:dirac_coord_point}
\end{equation}
where $V_\text{C} (r) = - \alpha Z/r$ is the Coulomb field of the nucleus. The wavefunction $\Psi (\boldsymbol{r})$ can be represented as
\begin{equation}
\Psi (\boldsymbol{r}) = \frac{1}{r}{G(r) \Omega_{jlm}(\hat{\boldsymbol{r}}) \choose i F(r) \Omega_{j\overline{l}m}(\hat{\boldsymbol{r}})},
\label{eq:psi_coord}
\end{equation}
where $\Omega_{jlm}$ is the spin spherical harmonic, $\overline{l} = 2j - l$, and $\hat{\boldsymbol{r}} \equiv \boldsymbol{r}/|\boldsymbol{r}|$. This leads to the radial equations:
\begin{eqnarray}
G^\prime + \frac{\kappa}{r}G - (E + m - V_\text{C})F = 0, \label{eq:FG1} \\
F^\prime - \frac{\kappa}{r}F + (E - m - V_\text{C})G = 0, \label{eq:FG2}
\end{eqnarray}
where $\kappa = \pm (j + 1/2)$ for $j = l \mp 1/2$ . These equations can be solved either analytically (see, e.~g.,~Ref.~\cite{blp}) or numerically.

\indent We will consider the nuclear recoil effect within the Breit approximation. For the point-like nucleus, it can be described by the operator \cite{shabaev1985}:
\begin{equation}
H_\text{B} = \frac{\boldsymbol{p}^2}{2M} - \frac{\alpha Z}{2Mr} \bigg ( \boldsymbol{\alpha} + \frac{\boldsymbol{\alpha} \cdot \boldsymbol{r}}{r^2} \, \boldsymbol{r} \bigg) \cdot \boldsymbol{p}.
\label{eq:breit}
\end{equation}
The full relativistic theory of the recoil effect is much more complicated and requires using QED beyond the Breit approximation~\cite{shabaev1985, pachucki_grotch, shabaev1998, adkins_2007}. If the operator~(\ref{eq:breit}) is included in the Dirac Hamiltonian, Eqs.~(\ref{eq:FG1}) and (\ref{eq:FG2}) have additional terms. The following radial equations can be easily obtained:
\begin{eqnarray}
\Big ( G' + \frac{\kappa}{r}G \Big ) - \Big ( E + m - V_\text{C}\Big ) F - \frac{1}{2M} \bigg [F'' - \frac{\kappa (\kappa - 1)}{r^2} F \bigg ] -  \frac{\alpha Z}{2Mr} \bigg [ 2G' + \bigg ( \frac{\kappa}{r} - \frac{1}{r} \bigg ) G \bigg ]&=& 0,\quad \label{eq:FG1_recoil_breit} \\
\Big ( F' - \frac{\kappa}{r}F \Big ) + \Big ( E - m - V_\text{C} \Big ) G + \frac{1}{2M} \bigg [G'' - \frac{\kappa (\kappa + 1)}{r^2} G \bigg ] -  \frac{\alpha Z}{2Mr} \bigg [ 2F' - \bigg( \frac{\kappa}{r} + \frac{1}{r} \bigg ) F \bigg ]&=& 0.\quad \label{eq:FG2_recoil_breit}
\end{eqnarray}
For the $1s$-state ($l = 0$, $\overline{l} = 1$, $\kappa = -1$) this reads
\begin{eqnarray}
\Big ( G' - \frac{1}{r}G \Big ) - \Big ( E + m - V_\text{C} \Big ) F - \frac{1}{2M} \bigg [F'' - \frac{2}{r^2} F \bigg ] -  \frac{\alpha Z}{Mr} \bigg [ G' - \frac{1}{r} G \bigg ]&=& 0, \label{eq:FG1_recoil_breit_1s} \\
\Big ( F' + \frac{1}{r}F \Big ) + \Big ( E - m - V_\text{C} \Big ) G + \frac{1}{2M} G''  -  \frac{\alpha Z}{Mr} F' &=& 0. \label{eq:FG2_recoil_breit_1s}
\end{eqnarray}
The presence of the second derivatives and terms $\sim 1/r^2$ considerably alters the asymptotic behavior of solutions in the vicinity of the nucleus (see Section~\ref{sec:results}).

\indent We will also examine the vacuum-polarization effect. The leading contribution of this effect is described by the Uehling potential which can be included in the Dirac equation nonperturbatively (in terms of ordinary quantum mechanics). This allows us to analyze the principle behavior of the solutions whereas methods based on the calculations of the expectation values can only provide certain corrections to the energy eigenvalues.

\indent The Uehling potential for the case of a point nucleus can be represented as (see, e.~g.,~Ref.~\cite{blp})
\begin{equation}
V_\text{U}(r) = - \frac{2\alpha (\alpha Z)}{3 \pi r} \int \limits_1^\infty \mathrm{d}t \frac{\sqrt{t^2 - 1}}{t^2} \Big ( 1 + \frac{1}{2t^2} \Big )\mathrm{e}^{-2mrt}.
\label{eq:uehling_coord}
\end{equation}
It is well-known that this operator decreases the energy value so it is unlikely to ``regularize'' the Dirac Hamiltonian for a point-nucleus model. Indeed, it turnes out that the Hamiltonian including the Uehling potential provides the $1s$ solution for the case $Z > 137$ only when it also contains the nuclear recoil operator. Furthemore, the Uehling potential should be analyzed only for the case of an extended nucleus since its asymptotic expansion for $r \to 0$ forbids any regular solutions ($V_\text{U} (r) \sim \ln r /r$ if $r \to 0$). However, the energy can be found if we consider an extended-nucleus model decreasing the nuclear radius $R$. In this case the Uehling potential has a finite value for $r = 0$, e.~g., for a homogeneously charged sphere with radius $R = \sqrt{5/3}\, \langle r^2 \rangle^{1/2}$:
\begin{equation}
V_\text{U}(0) = - \frac{\alpha (\alpha Z)}{\pi m R^3} \int \limits_1^\infty \mathrm{d}t \frac{\sqrt{t^2 - 1}}{t^3} \Big ( 1 + \frac{1}{2t^2} \Big ) \Big ( \frac{1}{2mt} - R\mathrm{e}^{-2mRt} - \frac{1}{2mt}\mathrm{e}^{-2mRt} \Big ).
\label{eq:uehling_zero}
\end{equation}

\indent The Dirac equation including the nuclear recoil and vacuum-polarization operators was solved numerically. The boundary conditions were obtained from the asymptotic expansions given in Appendix. The results are presented in the next section.
%%%
\section{Results}
\label{sec:results}
\begin{figure}[h]
\center{\includegraphics[width=0.8\linewidth]{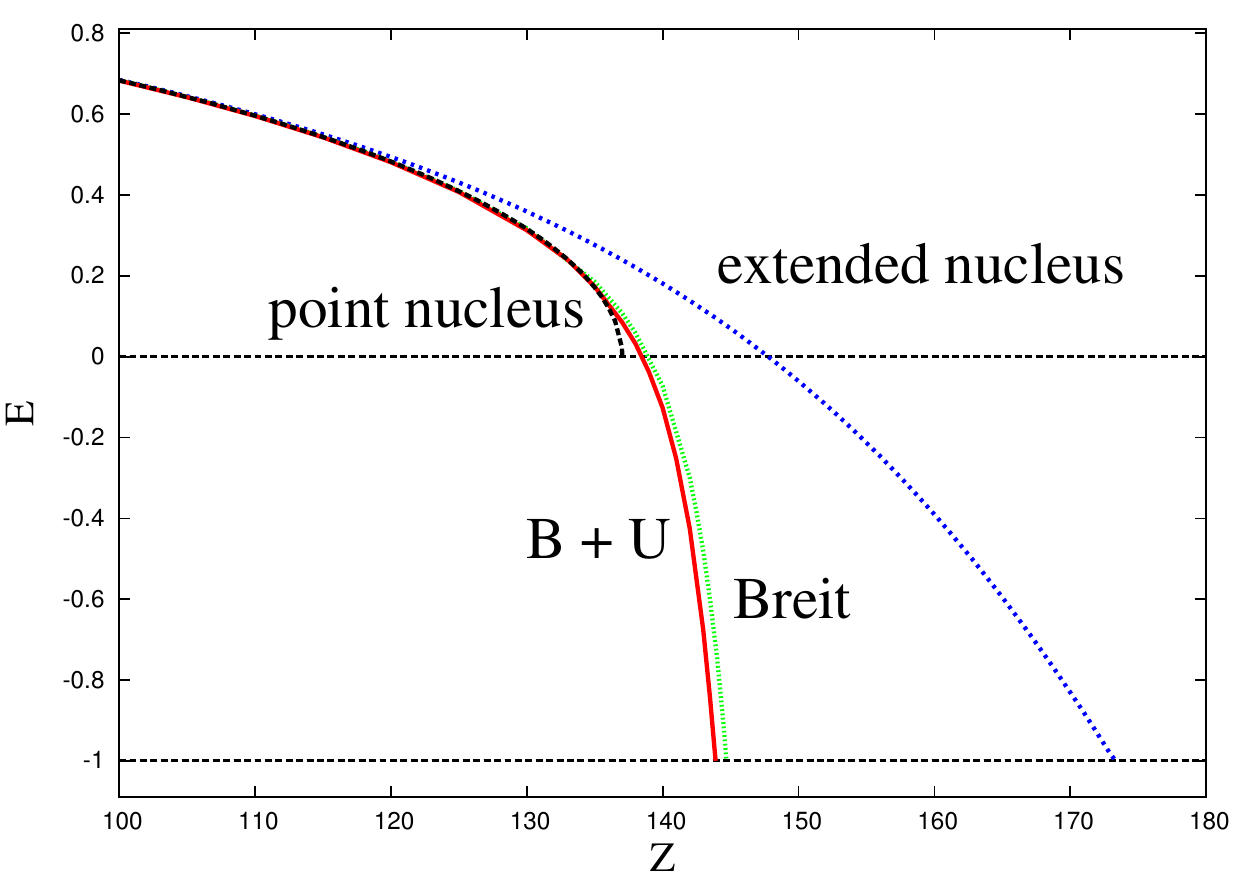}}
\caption{The $1s$-state energy evaluated according to the Sommerfeld formula (line ``point nucleus''), calculated for the case of an extended nucleus (line ``extended nucleus''), for the point-nucleus model with the nuclear recoil operator (line ``Breit'') and for the point-nucleus model with the recoil operator and the Uehling potential included in the Dirac Hamiltonian (line~``B~+~U'').}
\label{fig:energy_Z}
\end{figure}

We conducted numerical calculations in order to analyze the contributions of the nuclear recoil and vacuum-polarization effects in more detail. The $1s$-state energy corresponding to the Hamiltonian which contains the nuclear recoil operator within the Breit approximation is presented in Fig.~\ref{fig:energy_Z}. We have assumed that the nuclear mass as a function of the nuclear charge is given by $M = 2.6 Z m_\text{p}$, where $m_\text{p}$ is the proton mass. The wave function was constructed even for $Z > 137$ whereas the similar procedures for the motionless nucleus do not provide any adequate solutions. This confirms the hypothesis that the nuclear recoil operator ``regularizes'' the Dirac Hamiltonian for the point-nucleus model. When the Hamiltonian contains both the recoil operator and the Uehling potential the $1s$-state energy reaches the negative energy continuum at $Z \approx 144$ (this corresponds to the line denoted by ``B + U'' in Fig.~\ref{fig:energy_Z}). Taking into account the effects mentioned makes this value greater than $137$. The $1s$-state energy for the model of a homogeniously charged nucleus is also displayed in Fig.~\ref{fig:energy_Z} (the nuclear radius is assumed to be $\langle r^2 \rangle^{1/2} = 1.2 \, (2.6 Z)^{1/3}~\text{fm}$). The results are also presented in Table~\ref{table:res}.
\begin{table} [t]
\centering
\setlength{\tabcolsep}{0.5em}
\begin{ruledtabular}
\begin{tabular}{c|lllll}
\multirow{2}*{$Z$}&
\multicolumn{4}{c}{$E_{1s}$}\\ &
\raisebox{0pt}[10pt][4pt]{Sommerfeld formula} &
\raisebox{0pt}[10pt][4pt]{Breit} &
\raisebox{0pt}[10pt][4pt]{Breit + Uehling} &
\raisebox{0pt}[10pt][4pt]{Ext. nucleus}\\
\hline
$100$ & $0.6837$ & $0.6837$ & $0.6834$ & $0.6850$\\
$110$ & $0.5964$ & $0.5964$ & $0.5957$ & $0.6000$\\
$120$ & $0.4829$ & $0.4829$ & $0.4815$ & $0.4943$\\
$130$ & $0.3163$ & $0.3176$ & $0.3128$ & $0.3592$\\
$135$ & $0.1717$ & $0.1856$ & $0.1734$ & $0.2766$\\
$137$ & $0.02292$ & $0.1083$ & $0.08621$ & $0.2401$\\
$138$ & & $0.05911$ & $0.03228$ & $0.2210$\\
$140$ & & $-0.07288$ & $-0.1271$ & $0.1810$\\
$142$ & & $-0.2997$ & $-0.4249$ & $0.1386$\\
$143$ & & $-0.4818$ & $-0.6772$ & $0.1163$\\
$143.5$ & & $-0.6017$ & $-0.8450$ & $0.1049$\\
$144$ & & $-0.7435$ & & $0.09338$\\
$144.5$ & & $-0.9165$ & & $0.08164$
\end{tabular}
\end{ruledtabular}
\caption{The $1s$-state energy evaluated by using the Sommerfeld formula (second column), calculated for the point-nucleus model with the Hamiltonian containing only the nuclear recoil operator (third column) and both the recoil operator and the Uehling potential (fourth column), and calculated for the extended-nucleus model (last column). All values are in the relativistic units.}
\label{table:res}
\end{table}

\begin{figure}[h]
\center{\includegraphics[width=0.8\linewidth]{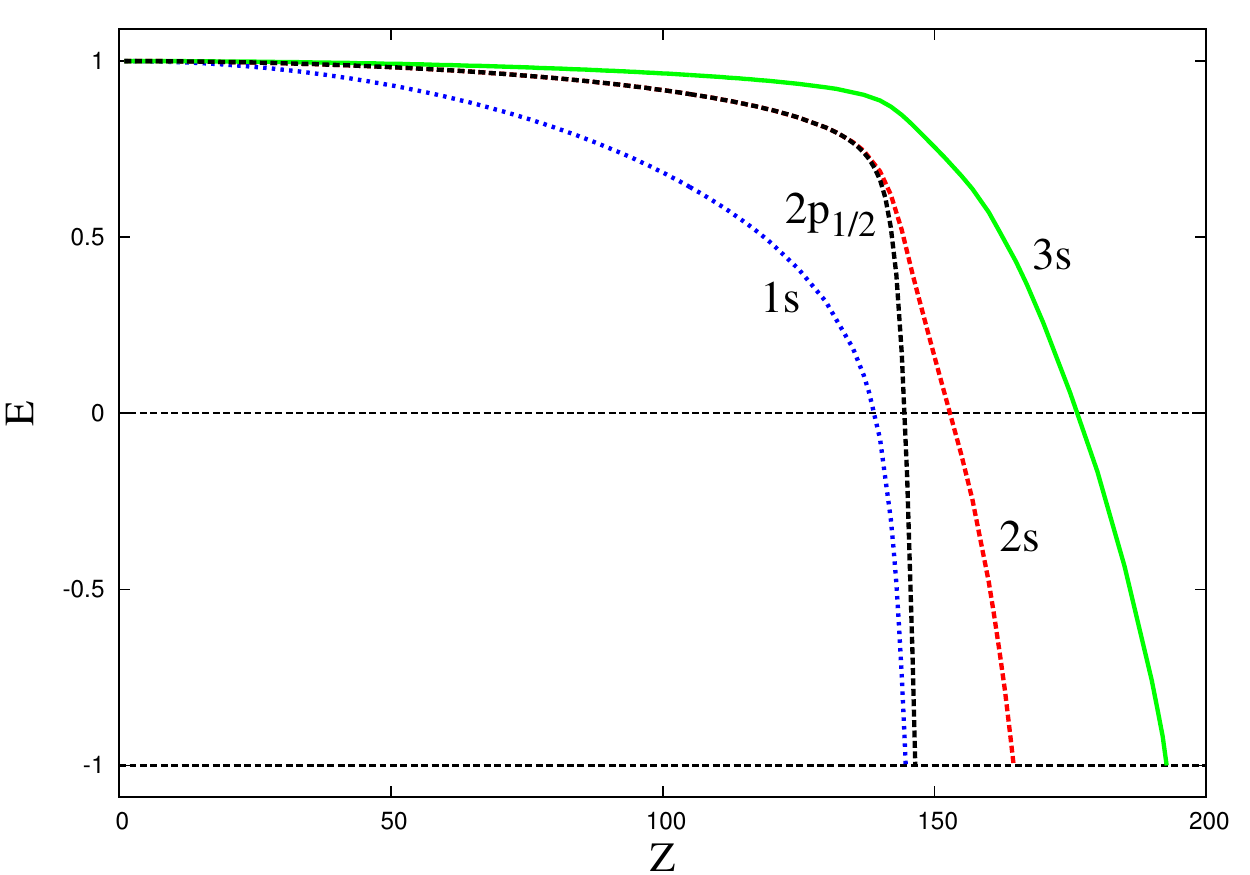}}
\caption{The energy of the $1s$, $2s$, $2p_{1/2}$ and $3s$ states calculated for the point-nucleus model with the nuclear recoil operator within the Breit approximation.}
\label{fig:energy_other_states}
\end{figure}

\indent In Ref.~\cite{gaertner_1981} it was shown that, when the nuclear radius tends to zero, all electronic states with $\alpha Z>|\kappa|$ one after the other reach the lower energy continuum that leads to the screening effect (this happens till the effective nuclear charge is $137$). This is due to the fact that for a point-like nucleus all $s$ and $p_{1/2}$ states have the critical charge $Z_\text{cr} \approx 137$. We also provided the calculations for $2s$, $2p_{1/2}$, $3s$ states and concluded that the nuclear recoil effect allows the nucleus to have an arbitrarily large effective charge since for any value of $Z$ the number of supercritical states is finite and, therefore, the nucleus is only partially screened by the vacuum charge. In Fig.~\ref{fig:energy_other_states} the energy as a function of $Z$ is presented for several $s$ and $p_{1/2}$ states. We see that for the $1s$, $2s$, $2p_{1/2}$, and $3s$ states $Z_\text{cr} \approx 145, 165, 146$, and $193$, respectively.
%
%%%
\section{Mathematical background}
\label{sec:math}
The first attempt to define the Dirac Hamiltonian as a self-adjoint operator in the presence of the Coulomb potential singularity was made in Ref.~\cite{case_1950}. More rigorous and comprehensive treatment of the problem was presented in Refs.~\cite{burnap_1981, xia_1999, hogreve_2013, gitman_2013}. It was shown that the Dirac Hamiltonian formally defined by Eq.~(\ref{eq:dirac_coord_point}) has a one-parameter family of self-adjoint extensions for $Z>137$ but none of them can be considered as the distinguished one. Since the defficiency indices of the symmetric (in $C_0^\infty (\mathbb{R}^3)$ which is dense in $L^2 (\mathbb{R}^3)$) operator (\ref{eq:dirac_coord_point}) are $(1, 1)$, the self-adjoint extensions exist. The corresponding theorem can be found, for instance, in Ref.~\cite{voronov_russ_physj} (see also Ref.~\cite{fulton_2012}).

\indent When one adds the nuclear recoil operator~(\ref{eq:breit}) the Hamiltonian remains symmetric. The same methods can be invoked in order to define it as a self-adjoint operator but this would be beyond the scope of the present paper.
%%%
\section{Discussion and conclusion}
\label{sec:discussion}
In this paper a supercritical hydrogenlike ion was considered regarding the nuclear recoil and vacuum-polarization effects. It was proved that the Dirac Hamiltonian even for the point-nucleus model has the $1s$-state for $Z > 137$, provided the nuclear recoil operator is included. If we take into account the Uehling potental and the nuclear recoil effect within the Breit approximation, the energy of this state reaches the negative energy continuum at $Z_\text{cr} \approx 144$. This critical charge is not a fundamental value since it should alter when one takes into account the other QED and higher-order relativistic recoil corrections. However, this paper demonstrates that, in contrast to the previous findings~\cite{gaertner_1981}, the interaction with a point charge in QED can effectively have the coupling strength greater than~$1$. Moreover, the analysis of other electronic states indicated that the effective nuclear charge can be arbitrarily large.
%%%
\section*{Acknowledgments}
This work was supported by RFBR (Grant No.~13-02-00630) and by SPbSU (Grants No. 11.42.1478.2015, 11.38.269.2014, and 11.38.237.2015). I. A. A. acknowledges the support from the ``Dynasty'' foundation and from the German-Russian Interdisciplinary Science Center (G-RISC) funded by the German Federal Foreign Office via the German Academic Exchange Service (DAAD).
%%%
\appendix*
\section{Asymptotic behavior of solutions}
\label{sec:appendix}
Let us, first, consider the Dirac Hamiltonian for a motionless nucleus (see Eqs.~(\ref{eq:FG1}) and (\ref{eq:FG2})). For the irregular singular point $r = \infty$ the Laplace method can be employed (i.~e. we represent the solution as $G(r) = \mathrm{e}^{-\lambda r} r^{-\rho} \sum c_k r^{-k}$). This leads to the following asymptotic expansions:
\begin{eqnarray}
G(r) &=& A\sqrt{m + E} \, \mathrm{e}^{-\lambda r},~r \to \infty, \label{eq:G_asym_inf} \\
F(r) &=& -A\sqrt{m - E} \, \mathrm{e}^{-\lambda r},~r \to \infty, \label{eq:F_asym_inf}
\end{eqnarray}
where $\lambda = \sqrt{m^2 - E^2}$. In the vicinity of the origin the asymptotic behavior can be analyzed by using the Frobenius method since $r=0$ is a regular singular point. For the case of a point nucleus one can obtain:
\begin{eqnarray}
G(r) &=& B r^\gamma,~r \to 0, \label{eq:G_asym_zero} \\
F(r) &=& B \, \frac{\kappa + \gamma}{\alpha Z} \, r^\gamma,~r \to 0, \label{eq:F_asym_zero}
\end{eqnarray}
where $\gamma = \sqrt{\kappa^2 - (\alpha Z)^2}$.

\indent If the Hamiltonian contains the nuclear recoil operator defined by Eq.~(\ref{eq:breit}), i.~e. the nuclear recoil effect is taken into account within the Breit approximation, then the asymptotic expansions for $r \to \infty$ alter slightly:
\begin{eqnarray}
G(r) &=& A \, \mathrm{e}^{-\lambda r},~r \to \infty, \label{eq:G_asym_inf_recoil} \\
F(r) &=& -\frac{A}{\lambda} \bigg ( m - E - \frac{\lambda^2}{2M} \bigg ) \mathrm{e}^{-\lambda r},~r \to \infty, \label{eq:F_asym_inf_recoil}
\end{eqnarray}
where $\lambda = \sqrt{2M \big [\sqrt{M^2 + 2ME + m^2} - (M+E) \big ]}$. Note, that $\lambda \to \sqrt{m^2 - E^2}$ if $m/M \to 0$, so in this limit the expressions obtained coincide with (\ref{eq:G_asym_inf}) and~(\ref{eq:F_asym_inf}). However, for the point $r = 0$ the asymptotic behavior becomes strongly different. It is possible to construct the only solution of Eqs.~(\ref{eq:FG1_recoil_breit_1s}) and~(\ref{eq:FG2_recoil_breit_1s}) which is holomorphic in the vicinity of the origin. Its asymptotic expansion does not depend on the fact weather $\alpha Z > 1$ or not:
\begin{eqnarray}
G(r) &=& Br + O(r^3),~r \to 0, \label{eq:G_asym_zero_breit} \\
F(r) &=& \frac{1}{2}MBr^2 + O(r^3),~r \to 0. \label{eq:F_asym_zero_breit}
\end{eqnarray}

\indent As was mentioned in Section~\ref{sec:theory}, we have to study the Uehling potential only for the extended-nucleus model. If the nucleus is assumed to be a homogeneously charged sphere with radius $R$, then the asymptotic expansions in the presence of the recoil operator within the Breit approximation and the Uehling potential have the form:
\begin{eqnarray}
G(r) &=& Br + O(r^3),~r \to 0, \label{eq:G_asym_zero_breit_uehl} \\
F(r) &=& \frac{1}{3}M\frac{\alpha Z}{1 + (\alpha Z)^2} \big ( E - m - V_\text{C} (0) - V_\text{U} (0)\big )Br^3 + O(r^4),~r \to 0, \label{eq:F_asym_zero_breit_uehl}
\end{eqnarray}
where $V_\text{C} (0)$ and $V_\text{U} (0)$ are the Coulomb potential of the sphere and the Uehling potential evaluated at the origin: $V_\text{C} (0) = -3\alpha Z /2R$ and $V_\text{U} (0)$ is given by Eq.~(\ref{eq:uehling_zero}). We conduct these calculations for~$R \to 0$.
%%%%%

\end{document}